\documentclass[twoside,twocolumn,9pt]{article}
\usepackage{extsizes}
\usepackage[super,sort&compress,comma]{natbib}
\usepackage[version=3]{mhchem}
\usepackage[left=1.5cm, right=1.5cm, top=1.785cm, bottom=2.0cm]{geometry}
\usepackage{balance}
\usepackage{times,mathptmx}
\usepackage{sectsty}
\usepackage{graphicx}
\usepackage{lastpage}
\usepackage[format=plain,justification=raggedright,singlelinecheck=false,font={stretch=1.125,small,sf},labelfont=bf,labelsep=space]{caption}
\usepackage{float}
\usepackage{fancyhdr}
\usepackage{fnpos}
\usepackage[english]{babel}
\usepackage{array}
\usepackage{droidsans}
\usepackage{charter}
\usepackage[T1]{fontenc}
\usepackage[usenames,dvipsnames]{xcolor}
\usepackage{setspace}
\usepackage[compact]{titlesec}
\usepackage{textcomp}
\usepackage{multirow}
\usepackage{amssymb}
\usepackage{amsmath}
\usepackage{bm}
\usepackage{wasysym}
\usepackage{hyperref} 

\newcommand{\onlinecite}[1]{\hspace{-1.0ex} \nocite{#1}\citenum{#1}}
\newcommand{\RW}[1]{\textcolor{black}{#1}}

\graphicspath{{./figures/}}

\definecolor{cream}{RGB}{222,217,201}

\begin{document}

\pagestyle{fancy}
\thispagestyle{plain}
\fancypagestyle{plain}{

\renewcommand{\headrulewidth}{0pt}
}

\makeFNbottom
\makeatletter
\renewcommand\LARGE{\@setfontsize\LARGE{15pt}{17}}
\renewcommand\Large{\@setfontsize\Large{12pt}{14}}
\renewcommand\large{\@setfontsize\large{10pt}{12}}
\renewcommand\footnotesize{\@setfontsize\footnotesize{7pt}{10}}
\makeatother

\renewcommand{\thefootnote}{\fnsymbol{footnote}}
\renewcommand\footnoterule{\vspace*{1pt}%
\color{cream}\hrule width 3.5in height 0.4pt \color{black}\vspace*{5pt}}
\setcounter{secnumdepth}{5}

\makeatletter
\renewcommand\@biblabel[1]{#1}
\renewcommand\@makefntext[1]%
{\noindent\makebox[0pt][r]{\@thefnmark\,}#1}
\makeatother
\renewcommand{\figurename}{\small{Fig.}~}
\sectionfont{\sffamily\Large}
\subsectionfont{\normalsize}
\subsubsectionfont{\bf}
\setstretch{1.125} 
\setlength{\skip\footins}{0.8cm}
\setlength{\footnotesep}{0.25cm}
\setlength{\jot}{10pt}
\titlespacing*{\section}{0pt}{4pt}{4pt}
\titlespacing*{\subsection}{0pt}{15pt}{1pt}

\fancyfoot{}
\fancyfoot[RO]{\footnotesize{\sffamily{1--\pageref{LastPage} ~\textbar  \hspace{2pt}\thepage}}}
\fancyfoot[LE]{\footnotesize{\sffamily{\thepage~\textbar\hspace{3.45cm} 1--\pageref{LastPage}}}}
\fancyhead{}
\renewcommand{\headrulewidth}{0pt}
\renewcommand{\footrulewidth}{0pt}
\setlength{\arrayrulewidth}{1pt}
\setlength{\columnsep}{6.5mm}
\setlength\bibsep{1pt}

\makeatletter
\newlength{\figrulesep}
\setlength{\figrulesep}{0.5\textfloatsep}

\newcommand{\topfigrule}{\vspace*{-1pt}%
\noindent{\color{cream}\rule[-\figrulesep]{\columnwidth}{1.5pt}} }

\newcommand{\botfigrule}{\vspace*{-2pt}%
\noindent{\color{cream}\rule[\figrulesep]{\columnwidth}{1.5pt}} }

\newcommand{\dblfigrule}{\vspace*{-1pt}%
\noindent{\color{cream}\rule[-\figrulesep]{\textwidth}{1.5pt}} }

\makeatother

\twocolumn[
  \begin{@twocolumnfalse}
\vspace{3cm}
\sffamily
\begin{tabular}{m{4.5cm} p{13.5cm} }

 & \noindent\LARGE{\textbf{Modelling the Mechanics and Hydrodynamics of Swimming {\em E. coli}}} \\
\vspace{0.3cm} & \vspace{0.3cm} \\

 & \noindent\large{Jinglei Hu,\textit{$^{a,b}$}  Mingcheng Yang,\textit{$^{c}$} Gerhard Gompper,\textit{$^{a}$} and Roland G. Winkler\textit{$^{a}$}} \\
 & \\
 & \noindent\normalsize{The swimming properties of an {\em E. coli}-type model bacterium are investigated by mesoscale hydrodynamic simulations, combining molecular dynamics simulations of the bacterium with the multiparticle particle collision dynamics method for the embedding fluid. The bacterium is composed of a spherocylindrical body with attached helical flagella, built up from discrete particles for an efficient coupling with the fluid. We measure the hydrodynamic friction coefficients of the bacterium and find quantitative agreement with experimental results of swimming {\em E. coli}. The flow field of the bacterium shows a force-dipole-like pattern in the swimming plane and \RW{two vortices perpendicular} to its swimming direction arising from counterrotation of the cell body and the flagella. By comparison with the flow field of a force dipole and rotlet dipole, we extract the force-dipole and rotlet-dipole strengths for the bacterium and find that counterrotation of the cell body and the flagella is essential for describing the near-field hydrodynamics of the bacterium.} \\
\end{tabular}

 \end{@twocolumnfalse} \vspace{0.6cm}

]

\renewcommand*\rmdefault{bch}\normalfont\upshape
\rmfamily
\section*{}
\vspace{-1cm}


\footnotetext{\textit{$^{a}$~Theoretical Soft Matter and Biophysics, Institute for
Advanced Simulation and Institute of Complex Systems,
Forschungszentrum J\"ulich, D-52425 J\"ulich, Germany; Email:  j.hu@fz-juelich.de, g.gompper@fz-juelich.de, r.winkler@fz-juelich.de}}
\footnotetext{\textit{$^{b}$~Address after August 2015: Kuang Yaming Honors School, Nanjing University, 210023 Nanjing, China;
 E-mail: hujinglei@nju.edu.cn}}
\footnotetext{\textit{$^{c}$~Beijing National Laboratory for Condensed
Matter Physics and Key Laboratory of Soft Matter Physics, Institute
of Physics, Chinese Academy of Sciences, Beijing 100190, China; E-mail: mcyang@iphy.ac.cn}}


\section{Introduction}

The bacterium {\em E. coli} is an example of a widely-studied class of motile microorganisms that exploit multiple helical flagella for locomotion. \cite{berg:04} Each flagellum is propelled by a reversible rotary motor anchored in the cell's membrane.\cite{bren:77,berg:03,berg:04} When all the motors rotate in the same direction, the flagella form a bundle and the bacterium swims forward; the so-called  `run' phase. It is interrupted by short periods of `tumble' events, where the reversal of the motor-rotation direction of some flagella causes the associated flagella to leave the bundle, thereby inducing erratic rotation of the cell body.\cite{berg:04,macn:77,Turner:2000jb,scha:02,Darnton:2007bj,kear:10} When the reversed motors switch back to their initial rotation direction, the bundle is reformed and the bacterium swims in a new direction. The alternating runs and tumbles allow the bacterium to efficiently execute a biased random walk toward favorable environments such as  food-concentrated regions by adjusting run and tumble durations to the environmental conditions.

Numerous experimental, theoretical, and simulation studies have been performed to unravel the physical aspects of bacteria locomotion.\cite{elge:15} These comprise aspects of the bacteria flagella such as their  polymorphic transformations \cite{call:75,macn:77,Turner:2000jb,wada:08,call:13,Vogel:2010epje,Vogel:2013prl} and bundle formation. \cite{reic:05,Darnton:2007jb, Graham:2011pre, Reigh:2012sm, Reigh:2013plosone,Larson:2010bj} Moreover,  bacterial propulsion properties have been investigated, \cite{Darnton:2007jb, Chattopadhyay:2006pnas,rode:13,flor:05,Larson:2010bj}  their run-and-tumble dynamics, \cite{Darnton:2007jb, Larson:2015sm} as well as the influence of hydrodynamic interactions on their motion adjacent to surface. \cite{Lauga:2006bj, Li:2008pnas,Shum:2010prsa, Leonardo:2011prl, Drescher:2011pnas,spag:12, Lemelle:2013sm,Stocker:2014prl,elge:15.1,Hu:2015scirep,shum:15} In addition, the effects of external flows on the dynamical behaviors of bacteria suspension have been addressed  \cite{Stocker:2012pnas,rusc:14,tour:14} along with their rheological properties. \cite{Rafai:2010prl,kaya:12,karm:14}

The complexity of the bundling and swimming processes, especially near-field hydrodynamics, poses substantial challenges for an analytical description of bacteria locomotion. Here, mesoscale hydrodynamics simulations are particularly valuable to gain insight into the microscopic aspects of swimming, because they are able to bridge the large length- and time-scale differences between the bacterium and fluid degrees of freedom.\cite{Kapral:2008acp,Gompper:2009aps} The multiparticle collision dynamics (MPC) method \cite{Kapral:1999jcp,Kapral:2008acp,Gompper:2009aps} has proven to be very valuable for the studies of active systems.\cite{ruec:07,goet:10,yang:11,kapr:08,elge:09,earl:07,elge:10,elge:13,thee:13,elge:15,Hu:2015scirep} Specifically, MPC has successfully been applied to elucidate synchronization between the flagella beating of nearby swimming sperm,\cite{yang:08} as well as bundling of helical  flagella of bacteria.\cite{Reigh:2012sm,Reigh:2013plosone}

Valuable theoretical insight into the swimming behavior of bacteria is achieved by very simplified models. An ellipsoidal or spherical body is often combined with either a single effective flagellum attached at one pole, \cite{Vogel:2010epje,voge:12,shum:15} or with several flagella attached in a more or less random manner. \cite{Larson:2010bj,Larson:2015sm,Hu:2015scirep} Thereby, hydrodynamic interactions are typically taken into  account by a hydrodynamic tensor,\cite{voge:12,shum:15,Larson:2010bj,Larson:2015sm} e.g., the Oseen or Rotne-Prager  tensor.\cite{dhon:96}  Such an approach is useful for a moderate number of hydrodynamically interacting units, typically a single or very few flagella. In contrast, studies of large systems and collective phenomena involving multi-flagellated cells require a different approach. We suggest to exploit the MPC method to account for fluid hydrodynamics. Moreover, a model for {\em E. coli}-type bacteria that can reproduce experimental results and allows for quantitative predictions of their swimming properties \RW{is desirable; here, only a few modelling studies are available so far.\cite{Larson:2010bj,Larson:2015sm,Hu:2015scirep}.}

In this article, we present a bacterium model which closely resembles the geometry, flagellar elastic properties, and rotary motor torque of {\em E. coli}. By MPC simulations, we show that this model quantitatively reproduces the experimentally measured properties of {\em E. coli} for both the hydrodynamic friction and the relation between the bacterial swimming speed and flagellar rotation rate. We find that the flow field near the model bacterium is rather complex with \RW{two} spiral vortices arising from counterrotation of the cell body and flagella. At larger distances, the flow field displays a dipole pattern comparable to that of swimming {\em E. coli}. We perform a detailed analysis of the flow by comparison with the force-dipole and rotlet-dipole approximations, which enables us to access the force-dipole and rotlet-dipole strengths, which are essential quantities for the description of bacterial swimming.

The rest of the manuscript is structured as follows. Section~\ref{sec:method_model} outlines the bacterium model and the MPC method for the fluid. Simulation results are presented in Sec.~\ref{sec:results}, and Sec.~\ref{sec:conclusion} summaries our findings. Further technical details for the flagellum modelling are presented in App. \ref{sec:app_model}, and the analysis of the flow field for periodic systems in App.~\ref{sec:app_force-dipol}, respectively.

\section{Simulation model and method} \label{sec:method_model}


\subsection{Bacterium model}

\begin{figure}[t!]
\centering
\includegraphics[width=1\columnwidth]{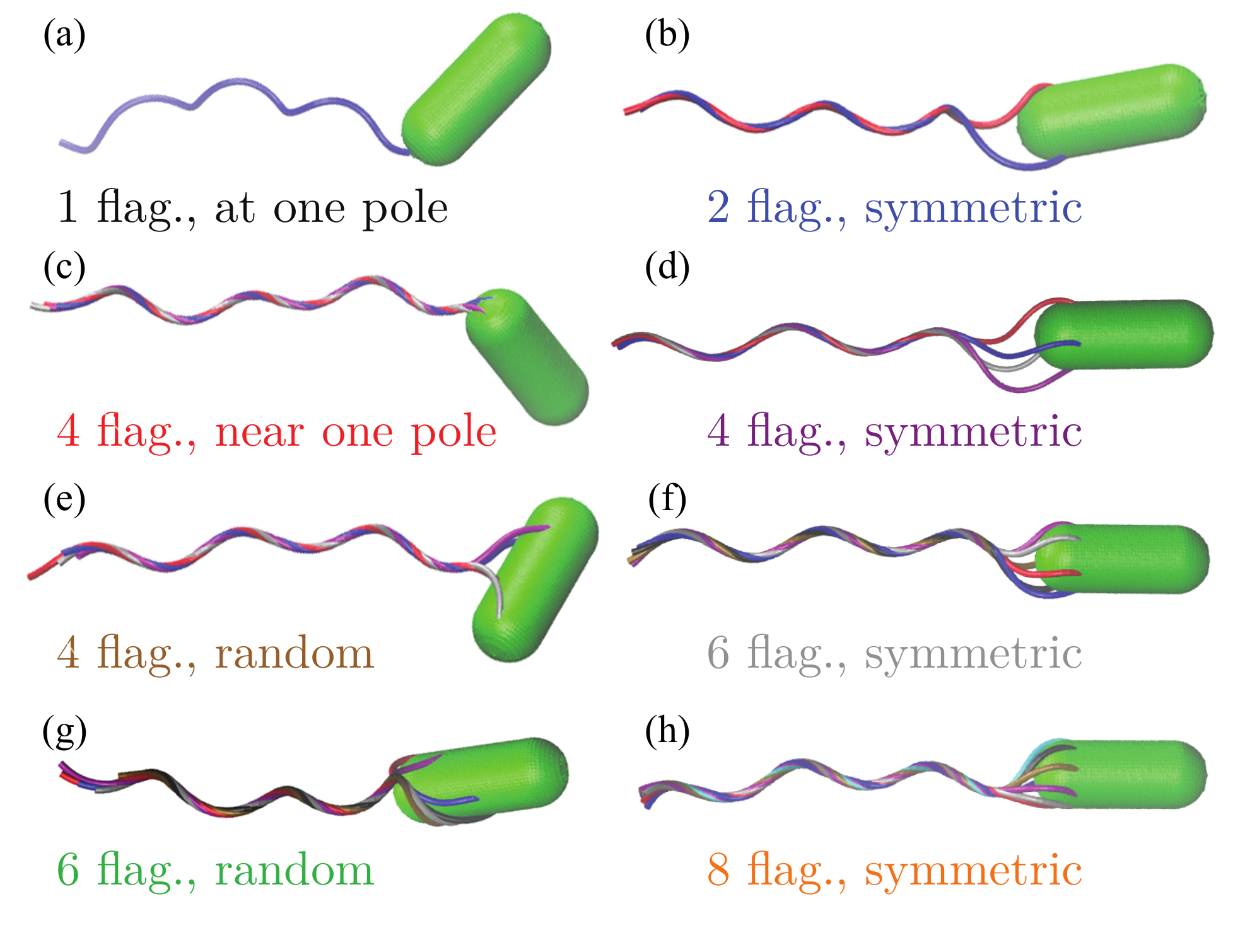}
\caption{Model of {\em E. coli}. Bacteria differ in the number of flagella and arrangement of flagella on the cell body. In a `symmetric' arrangement, the first contour particle of each flagellum is uniformly distributed along a circle on the body. In a `random' arrangement, the first contour particle is randomly located on the body. }
\label{Fig:1_1}
\end{figure}

\begin{figure}[h!]
\centering
\includegraphics[width=1\columnwidth]{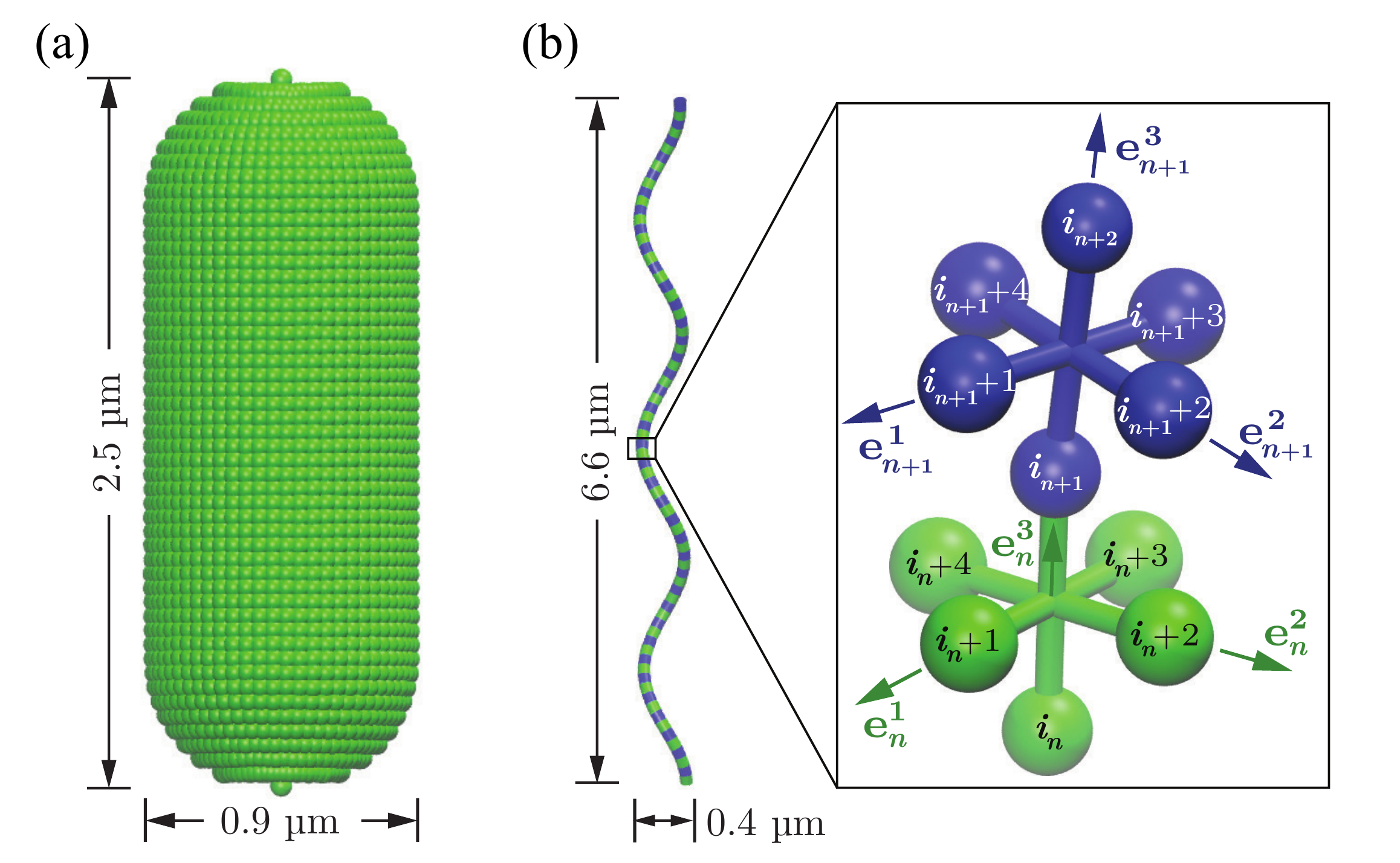}
\caption{(a) Model of the spherocylindrical cell body of diameter $d = 0.9$ \textmu m and length $\ell_{\rm b} = 2.5$ \textmu m. It is composed of 51 circular sections of particles, which are connected by the bond potential of Eq.~(\ref{Eq:Vbd}).  (b) The flagellum, a three-turn left-handed helix of radius $R=0.2$ \textmu m, pitch $\Lambda = 2.2$ \textmu m and contour length $L_c=7.6$ \textmu m (corresponding to the parallel length $L_{\|}=6.6$ \textmu m), consists of 76 consecutive segments. }
\label{Fig:1_2}
\end{figure}

The bacterium consists of a body and flagellar filaments, as shown in Fig.~\ref{Fig:1_1}, which are composed of point-like particles of mass $M=10m$. The cell body is represented by a spherocylinder of diameter $d = 9 a$ and length $l_{\rm b} = 25 a$, composed of 51 circular sections of particles with a spacing of $0.5\,a$, see Fig.~\ref{Fig:1_2}(a). Here, $m$ and $a$ are mass and length units related to the MPC fluid, as described in Sec.~\ref{sec:mpc-fluid}. Both pole sections consist of a single particle only. In each of the other 49 sections, 60 particles are uniformly distributed along circles on the spherocylinder surface. To maintain the shape of the body, nearest- and next-nearest-neighboring pairs of particles are bonded by a harmonic potential
\begin{equation}
U_{\rm bd}=\dfrac{1}{2}\,K_{\rm bd}(r-r_{\rm e})^2 ,
\label{Eq:Vbd}
\end{equation}
where $r$ and $r_e$ are the distance between the pair and the preferred value, respectively. Moreover,  two particles separated by 10, 20, and 30 particles along the 49 circular sections are additionally connected via the potential (\ref{Eq:Vbd}) in order to obtain stable circles. The two pole particles are only bonded to their 60 nearest neighbors. The bond strength $K_{\rm bd} = 10^4$ $k_BT/a^2$ is chosen for all bonds, whereas the preferred length $r_{\rm e}$ of each bond is determined by the geometry. Here, $k_B$ is the Boltzmann constant and $T$ the temperature.

 A flagellum is described by the helical wormlike chain model, \cite{Yamakawa:1997, Vogel:2010epje} with an adaptation suitable for the combination with MPC. As shown in Fig.~\ref{Fig:1_2}(b), a helical flagellum consists of $N = 76$ segments with a total of 381 particles. In each segment, six particles are arranged in an octahedron of edge length $a/\sqrt{2}$, forming 12 bonds along the edges and three along the diagonals. The preferred bond lengths are $r_{\rm e} = a/\sqrt{2}$ for edges and $r_{\rm e} = a$ for diagonals. The octahedron construction allows for a straightforward description of the intrinsic twist of the flagellum and a coupling of the twist to the forces exerted by the MPC fluid.

The bonds ${\bf b}_n^3 = {\bf r}_{i_{n+1}} - {\bf r}_{i_n}$ with $n = 1,...,N$ specify the contour of the flagellum, and, together with ${\bf b}_n^1 = {\bf r}_{i_n+1} - {\bf r}_{i_n+3}$ and ${\bf b}_n^2 = {\bf r}_{i_n+2} - {\bf r}_{i_n+4}$, define orthonormal triads $\{ {\bf e}_n^1,\,{\bf e}_n^2,\,{\bf e}_n^3\}$, where ${\bf e}_n^\alpha = {\bf b}_n^\alpha/\vert{\bf b}_n^\alpha\vert$ with $\alpha \in \{1, 2, 3\}$. Here, the ${\bf r}_{i_{n}}$  denote the positions of the backbone particles, and the ${\bf r}_{i_{n}+k}$ ($k =1,\ 2,\ 3,\ 4$) the positions of the particles in the plane with the normal  ${\bf e}_n^3$.

To characterize the local elastic deformation of a flagellum, the triad $\{ {\bf e}_n^1,\, {\bf e}_n^2,\, {\bf e}_n^3 \}$ is transported to $\{ {\bf e}_{n+1}^1,\, {\bf e}_{n+1}^2,\, {\bf e}_{n+1}^3 \}$ along the chain. This process is performed in two steps: (i) the rotation of $\{ {\bf e}_n^1,\, {\bf e}_n^2,\, {\bf e}_n^3 \}$ around ${\bf e}_n^3$ by a {\it twist} angle $\varphi_n$, and (ii) the rotation of the twisted triad $\{ {\tilde{\bf e}}_n^1,\, {\tilde{\bf e}}_n^2,\, {\tilde{\bf e}}_n^3 \}$ by a {\it bending} angle $\vartheta_n$ around the normal ${\bf n}_n = ({\bf e}_n^3 \times {\bf e}_{n+1}^3) / \vert {\bf e}_n^3 \times {\bf e}_{n+1}^3 \vert$ to the plane defined by the contour bonds ${\bf b}_n^3$ and ${\bf b}_{n+1}^3$. The elastic deformation energy is then
\begin{equation}
U_{\rm el} = \dfrac{1}{2}\sum_{\alpha=1}^3K_{\rm el}^\alpha\sum_{n=1}^{N-1}(\Omega_n^\alpha - \Omega_{\rm e}^\alpha)^2,
\label{Eq:Vel}
\end{equation}
where $K_{\rm el}^1=K_{\rm el}^2$ is the bending strength, $K_{\rm el}^3$ the twist strength, and  ${\bm{\Omega}}_n=\Omega_n^1 {\bf e}_n^1 + \Omega_n^2 {\bf e}_n^2 + \Omega_n^3 {\bf e}_n^3 = \vartheta_n {\bf n}_n + \varphi_n {\bf e}_n^3$ the strain vector. We choose $K_{\rm el}^1=K_{\rm el}^2=K_{\rm el}^3=5\cdot 10^4\, k_BT$, corresponding to a bending stiffness of $2\cdot 10^{-23}$ N m$^2$ for flagellar filaments within the experimental range of about $10^{-24}-10^{-21}$ N m$^2$. The parameters $\Omega_{\rm e}^\alpha$ in Eq. (\ref{Eq:Vel}) define the equilibrium geometry of the model flagellum and are chosen to recover the shape of an {\em E. coli} flagellum in the normal state, {\it i.e.}, a three-turn left-handed helix of radius 0.2 \textmu m and pitch 2.2 \textmu m. \cite{Darnton:2007jb} See the Appendix \ref{sec:app_model} for details.

A flagellum is attached to the cell body by choosing a body particle as its first contour particle ($i_1$, see Fig.~\ref{Fig:1_2}(b) for the notation). The rotation of the flagellum is induced by a motor torque ${\bf T}$ decomposed into a force couple ${\bf F}$ and $-{\bf F}$ acting on particles $i_1+2$ and $i_1+4$ (${\bf T}={\bf b}_1^2\times {\bf F}$ with ${\bf F}$ antiparallel to ${\bf b}_1^1$), or equivalently $i_1+1$ and $i_1+3$ (${\bf T}={\bf b}_1^1\times {\bf F}$ with ${\bf F}$ parallel to ${\bf b}_1^2$). Hence, there is no net force on the bacterium. We consider $\vert {\bf T} \vert \le 1000\,k_BT \simeq 4100$ pN nm, smaller than the stall torque of approximately 4500 pN nm of the flagellar motor. \cite{Berry:1997pnas} An opposite torque $-{\bf T}$ is applied to the body to ensure that the bacterium is torque-free. We do not explicitly model the hook that connects a flagellum and the body, \RW{in contrast to the model of Ref.~\onlinecite{Larson:2010bj},}  but incorporate the physical features of the hook: (i) transmitting the motor torque to the flagellum for rotation, and (ii) provide the flagellum the freedom to adopt any orientation relative to the body.  To prevent a flagellum from crossing the cell body or another flagellum, we use the repulsive Lennard-Jones potential
\begin{equation}
U_{\rm LJ} = \left\{
\begin{array}{ll}
4\,\epsilon [(\sigma/r)^{12}-(\sigma/r)^6] + \epsilon, & r \le \sqrt[6]{2}\,\sigma\\
0, &\textnormal{otherwise}
\end{array}
\right.
\label{Eq:VLJ}
\end{equation}
with $\epsilon = k_BT$ to capture excluded-volume interactions. For flagellum-body repulsion, $r$ is the distance between a flagellar contour particle and a body particle, and $\sigma = 0.5\, a$ is equal to the section spacing of the body. For flagellum-flagellum repulsion, $r$ is the closest distance between contour bonds of two flagella, and $\sigma = 0.25\, a$ is set by the filament diameter around 25 nm.\cite{Darnton:2007jb}

The dynamics of the bacterium is determined by the forces resulting from the potentials in Eqs. (\ref{Eq:Vbd})--(\ref{Eq:VLJ}) and the forces for generating the torques ${\bf T}$ and $-{\bf T}$, and by the momentum exchange with MPC fluid via the collision rule described by Eq. (\ref{Eq:Anderson}) below.

\subsection{Fluid model: Multiparticle collision dynamics}  \label{sec:mpc-fluid}

The MPC fluid is modeled by a collection of point-like particles of mass $m$. Their dynamics proceeds by alternating streaming and collision steps. In the streaming step, the fluid particles move ballistically and the position ${\bf r}_i$ of particle $i$ with its velocity ${\bf v}_i$ is updated according to
\begin{equation}
{\bf r}_i(t+\Delta t)={\bf r}_i(t)+{\bf v}_i(t)\,\Delta t,
\end{equation}
where $\Delta t$ is the time interval between collisions. In the collision step, all particles are sorted into cubic cells of length $a$ and the velocity ${\bf v}_i$ of particle $i$ in cell $c$ is renewed via the collision rule \cite{Noguchi:2007epl}
\begin{align}
{\bf v}_i^\text{new}&={\bf v}_c+{\bf v}_i^\text{ran}-{\textstyle\sum}_{j\in c} m_j{\bf v}_j^\text{ran}/{\textstyle\sum}_{j\in c} m_j\nonumber \\
&+\bigl[\mathbf{I}^{-1}{\textstyle\sum}_{j\in c}m_j({\bf r}_j-{\bf r}_c)\times ({\bf v}_j-{\bf v}_j^\text{ran})\bigr]\times ({\bf r}_i-{\bf r}_c),\label{Eq:Anderson}
\end{align}
where ${\bf v}_c$ and ${\bf r}_c$ are the center-of-mass velocity and position of the particles in $c$, $m_j$ the mass of particle $j$ in $c$, ${\bf v}_j^\text{ran}$ a random velocity sampled from the Maxwell-Boltzmann distribution, and $\mathbf{I}$ the moment-of-inertia tensor of all particles in $c$. The collision rule (i) conserves both linear and angular momentum in each cubic cell, (ii) includes thermal fluctuations of the fluid, and (iii) maintains a constant temperature. To satisfy Galilean invariance, a random shift of the collision-cell grid is performed before each collision step. \cite{Ihle:2001pre}

\subsection{Parameters}

The length $a$ of a collision cell, the mass $m$ of a MPC particle, and the thermal energy $k_BT$ define the length, mass, and energy units in our simulations. Other units are easily derived, e.g.,  density $\rho_0 = a^{-3}$, velocity $v_0 = \sqrt{k_BT/m}$, time $t_0 = a\sqrt{m/k_BT}$, and shear viscosity $\eta_0 = \sqrt{mk_BT}/a^2$.

The simulations are performed in cubic boxes of length up to $L=250a$ with periodic boundary conditions. The largest system contains more than $1.5 \times 10^8$ fluid particles. We choose the collision time step $\Delta t = 0.05 t_0$ and the fluid density $\rho = 10 \rho_0$, leading to the fluid viscosity $\eta = 7.15 \eta_0$ and the Schmidt number $Sc = 20$, for which the fluid exhibits liquid-like dynamics.\cite{ripo:05} Newton's equations of motion for the bacterium model are integrated with the time step $\delta t = \Delta t / 25$ using the velocity-Verlet algorithm. The Reynolds number $Re=d \rho v/ \eta$ is within the range 0.01--0.2 for the simulated bacteria with body diameter $d = 9 a$ and swimming speed $v= 0.00075-0.0125 v_0$.

Matching the geometry of our model bacteria to {\em E. coli} with a body of length of 2.5 \textmu m and diameter 0.9 \textmu m \cite{Darnton:2007jb} leads to the physical length $a \simeq 0.1$ \textmu m of a collision cell. The thermal energy is $k_BT \simeq 4.1$ pN nm at $T = 300$ K. A comparison with the viscosity of water of approximately $9\times10^{-4}$ N s$/$m$^2$ with  $\eta=7.15 \, \eta_0$ of the MPC fluid gives the physical time scale $t_0 \simeq 30$ \textmu s.

\begin{figure*}[t!]
\centering{\includegraphics[width=1.65\columnwidth]{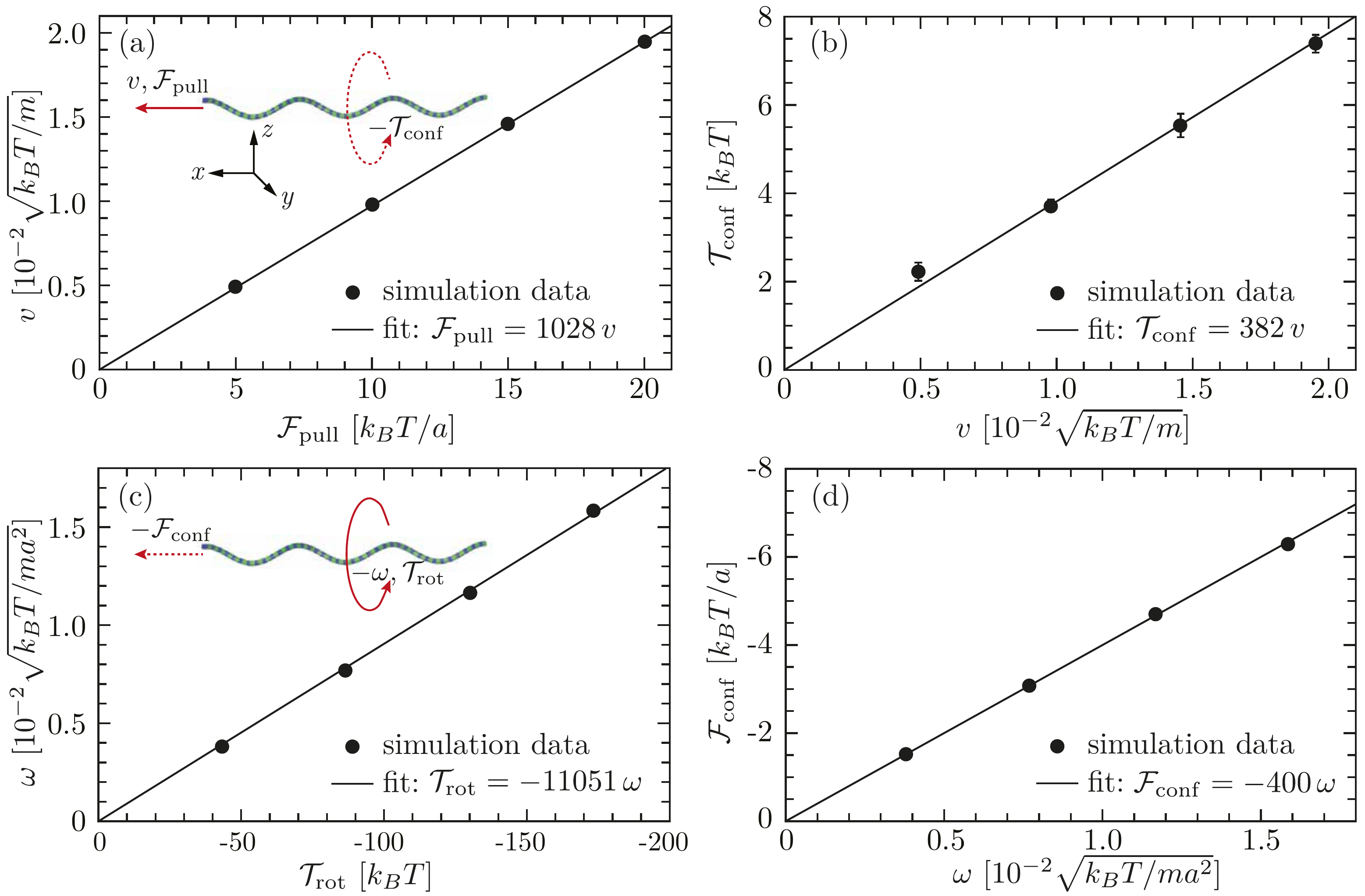}}
\caption{Results for a single flagellum from simulations in cubic boxes of length $L=12$ \textmu m: (a) translational speed $v$ versus axial force ${\cal F}_{\rm pull}$ and (b) confining torque ${\cal T}_{\rm conf}$ versus $v$ for a translating nonrotating flagellum; (c) rotation rate $\omega$ versus axial torque ${\cal T}_{\rm rot}$ and (d) confining force ${\cal F}_{\rm conf}$ versus $\omega$ for a rotating nontranslating flagellum. The red solid arrows indicate the direction of translation or rotation caused by ${\cal F}_{\rm pull}$ or ${\cal T}_{\rm rot}$, whereas the red dashed arrows imply the direction of rotation or translation which would have arisen from the rotation-translation coupling, but are prevented by ${\cal T}_{\rm conf}$ or ${\cal F}_{\rm conf}$ from the harmonic potentials on the flagellum.}
\label{Fig:2}
\end{figure*}

\section{Results and discussion} \label{sec:results}

\subsection{Hydrodynamic friction coefficients of the bacterium}
\noindent\subsubsection{Flagellum}

The flagellar hydrodynamic coefficients relate the force ${\cal F}$ and torque ${\cal T}$ exerted by the fluid on the flagellar bundle to its translational velocity $v$ and rotation rate $\omega$ (angular velocity $-\omega$ by right-hand rule) via
\begin{align}
{\cal F} &=  - \gamma_{\rm f}^{\rm t} v + \gamma_{\rm f}^{\rm c} (-\omega), \label{Eq:F}\\
{\cal T} &= +\gamma_{\rm f}^{\rm c} v -\gamma_{\rm f}^{\rm r} (-\omega) ,\label{Eq:T}
\end{align}
where $\gamma_{\rm f}^{\rm t}$ and $\gamma_{\rm f}^{\rm r}$ are the translational and rotational friction coefficients about the helical axis, and $\gamma_{\rm f}^{\rm c}$ is the rotation-translation coupling coefficient. These coefficients, known as elements of the propulsion matrix, \cite{Purcell:1997pnas} are essential quantities of bacterial swimming. We determine these coefficients by simulating a translating non-rotating ($\omega = 0$, $v \ne 0$) flagellum and a rotating non-translating ($v = 0$, $\omega \ne 0$) flagellum in the absence of the cell body. In the first case, the  flagellum is pulled by a force ${\cal F}_{\rm pull}$ along its helical axis, say $x$-axis, and $y$-, $z$-positions of each of its contour particles are trapped by harmonic potentials. This yields a confining torque ${\cal T}_{\rm conf}$ that prevents the rotation by counteracting the torque in Eq. (\ref{Eq:T}). In the second case, a flagellum is rotated by a motor torque, and harmonic potentials are applied on its contour particles that prevent its translation, but allow for the rotation about the helical axis, causing a net axial torque ${\cal T}_{\rm rot}$ and a confining force ${\cal F}_{\rm conf}$ opposite to the force in Eq. (\ref{Eq:F}). The two measurements lead to $\gamma^{\rm t}_{\rm f} = {\cal F}_{\rm pull} / v$, $\gamma^{\rm c}_{\rm f} = - {\cal T}_{\rm conf} / v$, $\gamma^{\rm r}_{\rm f} = -{\cal T}_{\rm rot} / \omega$, and $\gamma^{\rm c}_{\rm f} = {\cal F}_{\rm conf} / \omega$. In the latter measurement, two no-slip walls\cite{Lamura:2001epl, Luijten:2010jpcm} orthogonal to the flagellum  are included to ensure that  the flagellar speed $v$ vanishes with respect to the background fluid at rest. Without such no-slip walls, the translational speed of the flagellum is still $v$ relative to the fluid velocity at the boundaries since the rotating flagellum is constantly pumping fluid backwards. From the results shown in Fig.~\ref{Fig:2}, we obtain $\gamma^\text{t}_\text{f}/(m/t_0) = 1028 \pm 2$, $\gamma^\text{c}_\text{f}/\sqrt{m k_BT} = -382 \pm 7$, $\gamma^\text{r}_\text{f}/(k_BT t_0) = 11051 \pm 89$, and $\gamma^\text{c}_\text{f}/\sqrt{m k_BT} = -400 \pm 1$. The two estimates of $\gamma^\text{c}_\text{f}$ agree within 5\%, confirming the consistency of our measurements.

\noindent\subsubsection{Cell body}

In a similar way, we measure the translational and rotational friction coefficients $\gamma^{\rm t}_{\rm b}$ and $\gamma^{\rm r}_{\rm b}$ of the cell body by considering the body under an axial force ${\cal F}_{\rm pull}$ or torque ${\cal T}_{\rm rot}$. We obtain $\gamma^{\rm t}_{\rm b} = {\cal F}_{\rm pull} / v = (1118 \pm 5)m/t_0$ and $\gamma^{\rm r}_{\rm b} = {\cal T}_{\rm rot} / \omega = (52435 \pm 236) k_BT t_0$ from simulations with box size $L = $ 12 \textmu m. These values are consistent with the theoretical values $\gamma^{\rm t}_{\rm b}/(m/t_0)=1023$  and $\gamma^{\rm r}_{\rm b}/(k_BT t_0) = 46618$ for a solid spherocylinder of the same size.\cite{Norisuye:1979macromol,Yoshizaki:1980jcp} The quantitative agreement confirms that the applied particle-based mesoscale hydrodynamics model can well capture the no-slip boundary conditions on the bacterium surface, as already demonstrated for hard sphere colloids in Ref.~\citenum{pobl:14}.

\begin{table}[h!]
\caption{Comparison of hydrodynamic properties for the model bacterium with experimental results of {\em E. coli}. \cite{Chattopadhyay:2006pnas} $\gamma^{\rm t}$ and $\gamma^{\rm r}$ are the translational and rotational friction coefficients, and $\gamma^{\rm c}$ is the rotation-translation coupling coefficient. The subscripts `b' and `f' refer to body and flagellum, respectively. All the coefficients are rescaled by the absolute values of $\gamma^{\rm t}_{\rm b}$, which are 1281 and 1118 $m/t_0$ from simulations in cubic boxes of lengths $L=8$ and 12 \textmu m, and $\gamma^{\rm t}_{\rm b} = 1.4 \times 10^{-2}$ pN s$/$\textmu m from experiments.}\label{Tab:1}
\centering
\begin{tabular*}{0.48\textwidth}{@{\extracolsep{\fill}}llll}
\hline
\multirow{2}{*}{} & \multicolumn{2}{c}{model}  & \multirow{2}{*}{exp.} \\ \cline{2-3}
                         & $L=8$ \textmu m  & $L=12$ \textmu m & \\
\hline
$\gamma^{\rm r}_{\rm b}/\gamma^{\rm t}_{\rm b}$ [\textmu m$^2$] & 0.41 & 0.47  & 0.30  \\[3pt]
$\gamma^{\rm t}_{\rm f}/\gamma^{\rm t}_{\rm b}$ & 0.84  & 0.88  & 1.1  \\
$\gamma^{\rm r}_{\rm f}/\gamma^{\rm t}_{\rm b}$ [\textmu m$^2$] &  0.087  & 0.099  & 0.050  \\
$\gamma^{\rm c}_{\rm f}/\gamma^{\rm t}_{\rm b}$ [\textmu m]  & $-0.032$  & $-0.036$  & $-0.056$  \\
\hline
\end{tabular*}
\end{table}

\subsection{Comparison with experimental results and resistive-force theory}

Table \ref{Tab:1} summarizes the hydrodynamic coefficients of the bacterium body and a flagellum. The values measured for the system sizes $L=8$ and $12$ \textmu m differ by at most  10\%, implying that the periodic boundaries in the simulations have a rather small effect. The experimental values in Table~\ref{Tab:1} are from measurements of swimming {\em E. coli} \cite{Chattopadhyay:2006pnas} with an average body length of $3.0 \pm 0.8$ \textmu m, close to 2.5 \textmu m in our model. The flagellar properties determined from experiments are for a bundle with, on average,  $3-4$ flagella, rather than for a single flagellum. In addition, the cell body is found to wobble around the swimming axis. Given these caveats, the hydrodynamic properties of our model flagellum and cell body are in good agreement with experimental results for {\em E. coli}.

\begin{table}[h!]
\caption{Comparison of the friction coefficients of a flagellum with estimates by resistive-force theory (RTF) \cite{ligh:76} and experimental results of {\em E. coli}. \cite{Chattopadhyay:2006pnas} The theoretical expressions derived by resistive-force theory are presented in App.~\ref{sec:app_rft}. For the current model, only results for the box size $L=12$ \textmu m are presented. }\label{Tab:2}
\centering
\begin{tabular*}{0.48\textwidth}{@{\extracolsep{\fill}}llll}
\hline
  &  model & RTF & exp.\\
\hline
$|\gamma^{\rm c}_{\rm f}/\gamma^{\rm t}_{\rm f}|$ [\textmu m]     & 0.04  & 0.04 & 0.05  \\
$\gamma^{\rm r}_{\rm f}/\gamma^{\rm t}_{\rm f}$ [\textmu m$^2$] & 0.1  & 0.05  & 0.05  \\
$|\gamma^{\rm r}_{\rm f}/\gamma^{\rm c}_{\rm f}|$ [\textmu m] &  2.7 & 1.3  & 0.9  \\
\hline
\end{tabular*}
\end{table}

\RW{
Table \ref{Tab:2} compares the friction coefficients with results obtained by resistive-force theory (RTF) theory.\cite{ligh:76,Lauga:2006bj,Chattopadhyay:2006pnas,rode:13}  We use the theoretical expressions derived in  Ref.~\onlinecite{ligh:76}; the terms for the respective friction coefficients are summarized in  App.~\ref{sec:app_rft}. In order to avoid ambiguities by different expressions for the factors $K_n$ and $K_t$ of Eqs.~(\ref{eq_app_kn}) and (\ref{eq_app_trans}), as presented in Refs.~\onlinecite{ligh:76,Lauga:2006bj,Chattopadhyay:2006pnas}, we discuss ratios of the various friction coefficients only. For the theoretical results, we use the pitch angle $\zeta = 0.519$, the pitch length $\Lambda = 2.2$ \textmu m, and the hydrodynamic radius $r=0.1$ \textmu m of the flagellum string, which yields $\chi=0.68$ (cf. App.~\ref{sec:app_rft}). The ratio $\gamma^{\rm c}_{\rm f}/\gamma^{\rm t}_{\rm f}$ is in close agreement for the various approaches.  However, ratios for the simulation model including $\gamma^{\rm r}_{\rm f}$ are about twice larger than the prediction of RTF or  even three times compared to those determined from experimental results. As far as the comparison with RTF is concerned, the agreement is reasonable, considering the fact that the simulation model uses a discrete representation of the helix, and the ambiguities in the various parameters of the analytical approach. Here, a more thorough comparison between theory and simulation, and between the various theoretical approaches would be desirable. In experiments, the helical bundle seems to yield a much smaller rotational fiction coefficient. To which extend this friction coefficient is affected by the nature of the bundle and the above mentioned wobbling dynamics needs to be studied further.
}
\subsection{Swimming speed and flagellar rotation rate}

\begin{figure}[h!]
\centering{\includegraphics[width=\columnwidth]{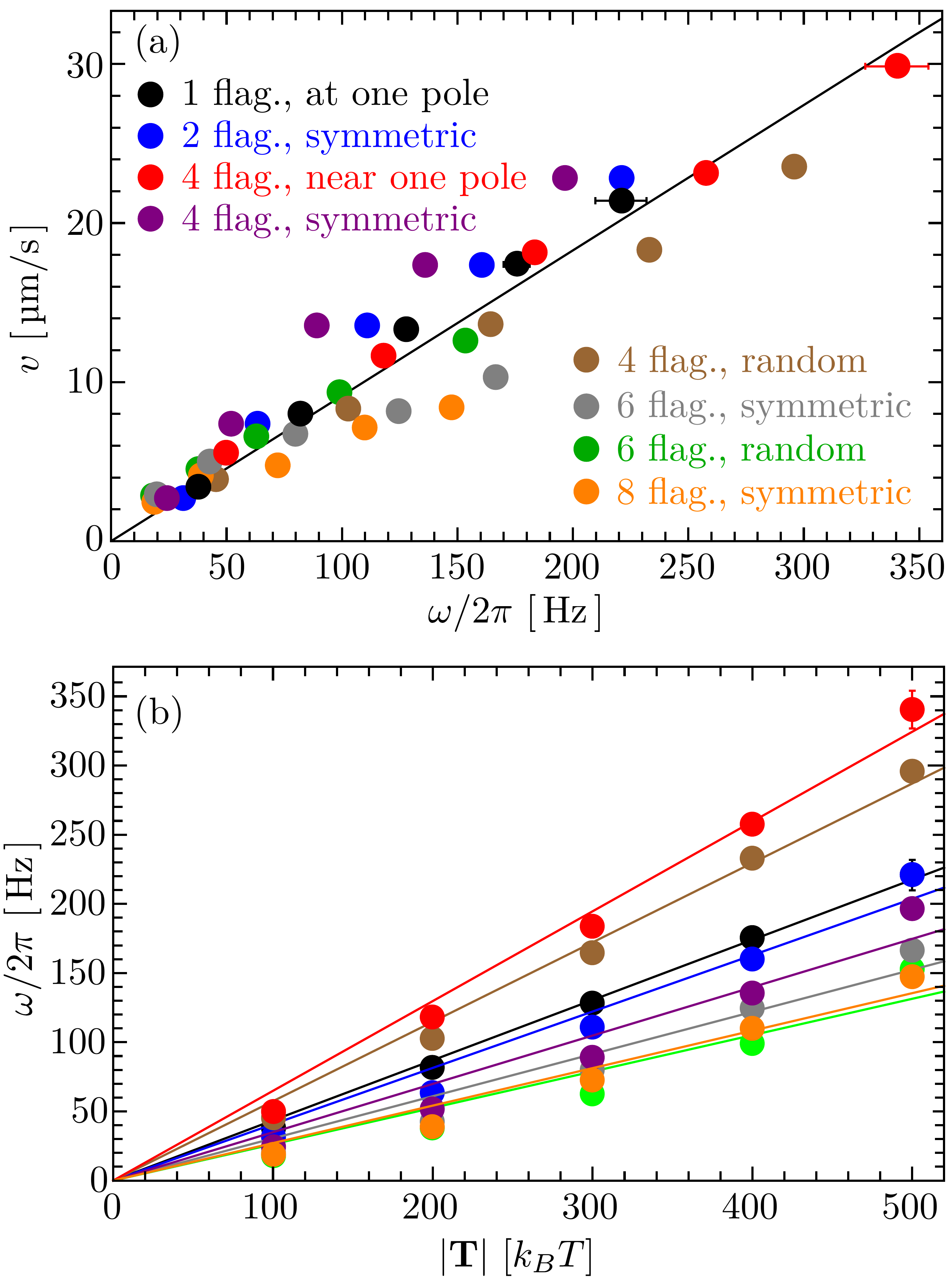}}
\caption{(a) Bacterial swimming speeds $v$ as function of the rotation frequency $\omega$ of the flagellar bundle  and (b) rotation frequency $\omega$ as function of  the flagellar motor torque $\vert {\bf T} \vert$ for the model bacteria shown in Fig.~\ref{Fig:1_1}. The solid line in (a) is a least-square fit to all data points. The solid lines in (b) are least-square fits to the data points in the same color.}
\label{Fig:3}
\end{figure}

In Fig.~\ref{Fig:3}(a), the bacterial swimming speed $v$ is displayed as a function of the rotation rate $\omega$ of the flagellar bundle for our model bacteria of Fig.~\ref{Fig:1_1}, which differ in the number of flagella and the arrangement of flagella on the cell body, ranging from one to eight flagella with symmetric and random arrangements of the anchoring points. The rotation rate of the flagellar bundle is $\omega = 2 \pi/\tau$, where $\tau$ is the average time for the helical bundle to complete one revolution. The data points in Fig.~\ref{Fig:3}(a) are obtained from extensive hydrodynamics simulations, where each bacterium travels a distance larger than its own length of about 10 \textmu m. The solid line in Fig.~\ref{Fig:3} is a least-square fit to all data points through the origin, yielding $v/\omega R=0.07$, comparable to the experimental ratio $v/\omega R = 0.14$, \cite{Darnton:2007jb} where $R = 0.2$ \textmu m is the helix radius of the {\em E. coli} flagellum. The scatter of the data points around the linear relation may arise from the fact that individual bacteria exhibit different wobbling amplitudes of the body.

The ratio of swimming speed $v$ to flagellar rotation rate $\omega$ can be understood from the hydrodynamic properties of the bacterium. The force in Eq. (\ref{Eq:F}) is balanced by the translational friction force $-\gamma^{\rm t}_{\rm b} v$ exerted by the fluid on the cell body, which gives
\begin{align}
\frac{v}{\omega} = -\frac{\gamma^{\rm c}_{\rm f}}{\gamma^{\rm t}_{\rm b}+\gamma^{\rm t}_{\rm f}} .
\end{align}
We obtain $v/\omega R \simeq 0.1$ and 0.13 using the model and experimental values in Table~\ref{Tab:1}, respectively. These ratios are close to $v/\omega R = 0.07$ from the fitted line in Fig.~\ref{Fig:3}(a) and the experimental ratio $v/\omega R = 0.14$. The agreement emphasizes the importance of hydrodynamic coefficients for a quantitative understanding of bacterial swimming.

\begin{figure*}[t!]
\centering{\includegraphics[width=2\columnwidth]{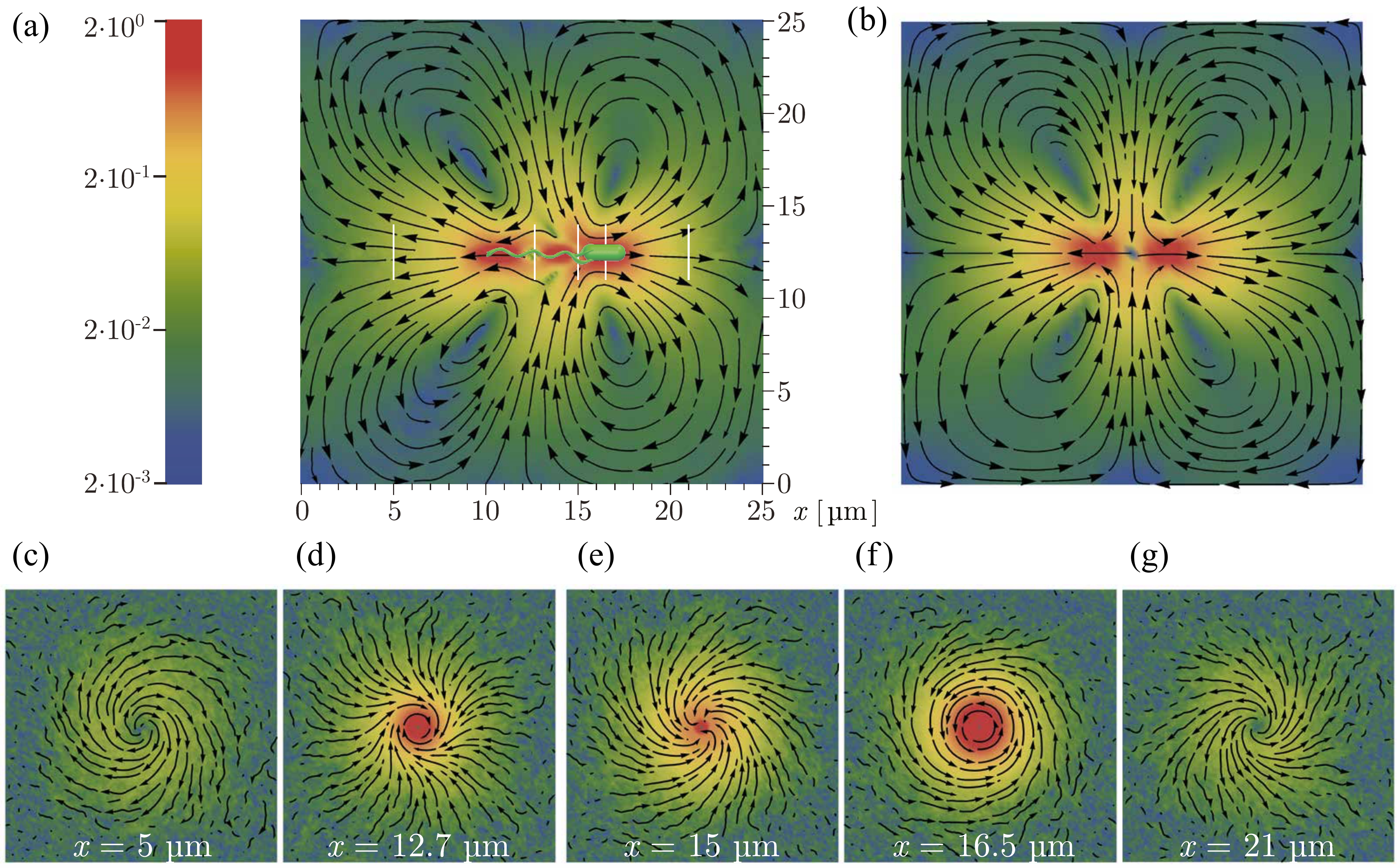}}
\caption{Time-averaged flow field generated by a single swimming bacterium as obtained from simulations: (a) flow field in the swimming plane \RW{(b)} the theoretical flow pattern for a finite-distance force dipole as illustrated in Fig.~\ref{Fig:5}(a) \RW{as superposition of two Stokeslets} within the same periodic box as for our simulations). (c) -(g) Flow fields in planes perpendicular to the swimming plane at positions indicated by the white vertical lines in (a). The streamlines indicate the flow direction, and the logarithmic color scheme indicates the magnitude of the flow speed scaled by the bacterial swimming velocity.}
\label{Fig:4}
\end{figure*}


Figure~\ref{Fig:3}(b) shows that the rotation rate of a bundle formed by four flagella near one pole of the cell body (red points) is larger than that of a single flagellum (black points) at the same motor torque, e.g., $\omega/2\pi = 340$ Hz versus 223 Hz at $\vert {\bf T} \vert = 500\, k_BT$. This result is consistent with the experimental observation \cite{Darnton:2007jb} that a flagellar bundle rotates faster than a single flagellum. The enhanced rotation as a result of the hydrodynamic coupling between the flagella and has also been found in simulations of helical bundles, \cite{Reigh:2012sm} where one of the ends of the helices is fixed in space. We note, however, that such hydrodynamic enhancement can be outweighed by the flagellum-body friction, depending on the connection of flagellar filaments to the body. As shown in Fig.~\ref{Fig:3}(b), the bundle of four flagella with a `symmetric' arrangement on the body (purple points, Fig.~\ref{Fig:1_1}(d)), rotates slower than both a single flagellum (black points, Fig.~\ref{Fig:1_1}(a)) and the bundle of four flagella near one pole of the body (red points, Fig.~\ref{Fig:1_1}(c)) at the same motor torque $\vert {\bf T} \vert \le 400 \, k_BT$.  In {\em E. coli}, each flagellum is attached to the body via a flexible hook of approximately 50 nm  length,  \cite{kawa:96} much shorter than the flagellar helix radius of 0.2 \textmu m. The rotating flagella that form a bundle spanning over the body are very likely to collide with the body, contributing to the flagellum-body friction. With the single-flagellum rotational friction coefficient $\gamma^{\rm r}_{\rm f} = 11051k_BT t_0$, the torque rotating the bundle of four flagella with $\omega / 2\pi = 196$ Hz ($\omega = 0.037/t_0$, see the purple point at $\vert {\bf T} \vert = 500\, k_BT$) is $\gamma^{\rm r}_{\rm f} \omega \simeq 409\, k_BT$, only about one-fifth of the total motor torque $2000\, k_BT$, indicating a significant flagellum-flagellum and flagellum-body friction.

\subsection{Flow field }

Figure \ref{Fig:4} shows the time-averaged flow field created by a model bacterium of approximately  8 \textmu m length in a periodic cubic \RW{simulation box} of length $L = 25$ \textmu m. Each of the four flagella is turned by a motor torque $\vert {\bf T}\vert = 1000$ $k_BT$, leading to a bundle with rotation frequency $\omega = 0.074/t_0$. The bacterial swimming speed is $v \simeq 0.0125 v_0$ and the propulsion force $f_{\rm p} = \gamma_{\rm b}^{\rm t} v \simeq 0.57$ pN consistent with the experimental values \RW{of} about 0.4--0.6 pN. \cite{Chattopadhyay:2006pnas, Drescher:2011pnas} Harmonic potentials are applied to $y$- and $z$-positions of the center-of-mass of the body and flagellar bundle such that the swimming axis is parallel to the $x$-axis. The flow field in the swimming plane, shown in Fig.~\ref{Fig:4}(a), is computed by averaging the velocities of the fluid particles cylindrically symmetrically around the swimming axis. The flow fields in Figs.~\ref{Fig:4}(c)--(g) are obtained \RW{as} time average of the fluid-particle velocities in the planes perpendicular to the swimming axis at \RW{the} different locations indicated by the white vertical lines in Fig.~\ref{Fig:4}(a).

The flow pattern not too close to the bacterium approximately resembles that of swimming {\em E. coli} determined from experiments (see Fig. 1A in Ref.~\citenum{Drescher:2011pnas}).  Closer to the bacterium, the flow field exhibits specific features reflecting the bacterium's detailed  structure. In particular, the flow field reveals a front-back asymmetry, since the cell body and flagellar bundle are physically different units. \RW{Along the cell axis, we find low fluid velocities in front of the cell, behind the cell body as well as in the middle and behind the flagellar bundle.  The flow velocity is high along the whole cell body, at the flagellar bundle somewhat behind the cell body, and toward the end of the bundle.}
The streamlines are closed in Fig.~\ref{Fig:4}(a) as a consequence of the applied periodic \RW{boundary conditions}, which implies differences in the far field compared to experimental observations. The effect of the boundary conditions is confirmed by the theoretical flow field for a finite-distance force dipole with the same boundary condition in Fig.~\ref{Fig:4}(b).


The flow patterns in the planes perpendicular to the swimming axis illustrate the interplay between the rotating flagellar bundle and counterrotating cell body. As shown in Figs.~\ref{Fig:4}(c), (e), and (g), the flow \RW{field} exhibits \RW{two spiral vortices associated with the rotation of the cell body and flagellar bundle, respectively.} In Fig.~\ref{Fig:4}(d), the fluid follows the clockwise rotation of the flagellar bundle in the central region, and tends to rotate counterclockwise in the outer region, implying a strong influence of the cell body. In Fig.~\ref{Fig:4}(f), the fluid rotates counterclockwise as the body and there is no significant effect from the flagellar rotation.

\RW{The flow field of the discrete-particle bacterium model of Ref.~\onlinecite{Larson:2010bj}, with a triangular-prism  body and three flagella, each composed of 15 beads, involves an infinite fluid domain, because hydrodynamic interactions are captured by the Rotne-Prager hydrodynamic tensor. Consistent with this study, we find spiral and helical flows (Figs.~\ref{Fig:4}(c)-(g)). However, there are also distinct differences, namely (i) three vortices are predicted in Fig.~5 of Ref.~\citenum{Larson:2010bj}, and (ii) fluid flows toward the flagellar bundle from behind (Figs. 4 and 7) rather than away from the cell as in Fig.~\ref{Fig:4}. It is not {\em a prior} evident where these differences come from. A reason could be the more detailed modelling of the cylindrical cell body and the flagella in our approach. Furthermore, in agreement with our studies, the experimentally determined flow field presented in Ref.~\onlinecite{Drescher:2011pnas} shows no indication of a forward flow at the end of the flagellar bundle. }

\begin{figure}[t!]
\centering{\includegraphics[width=\columnwidth]{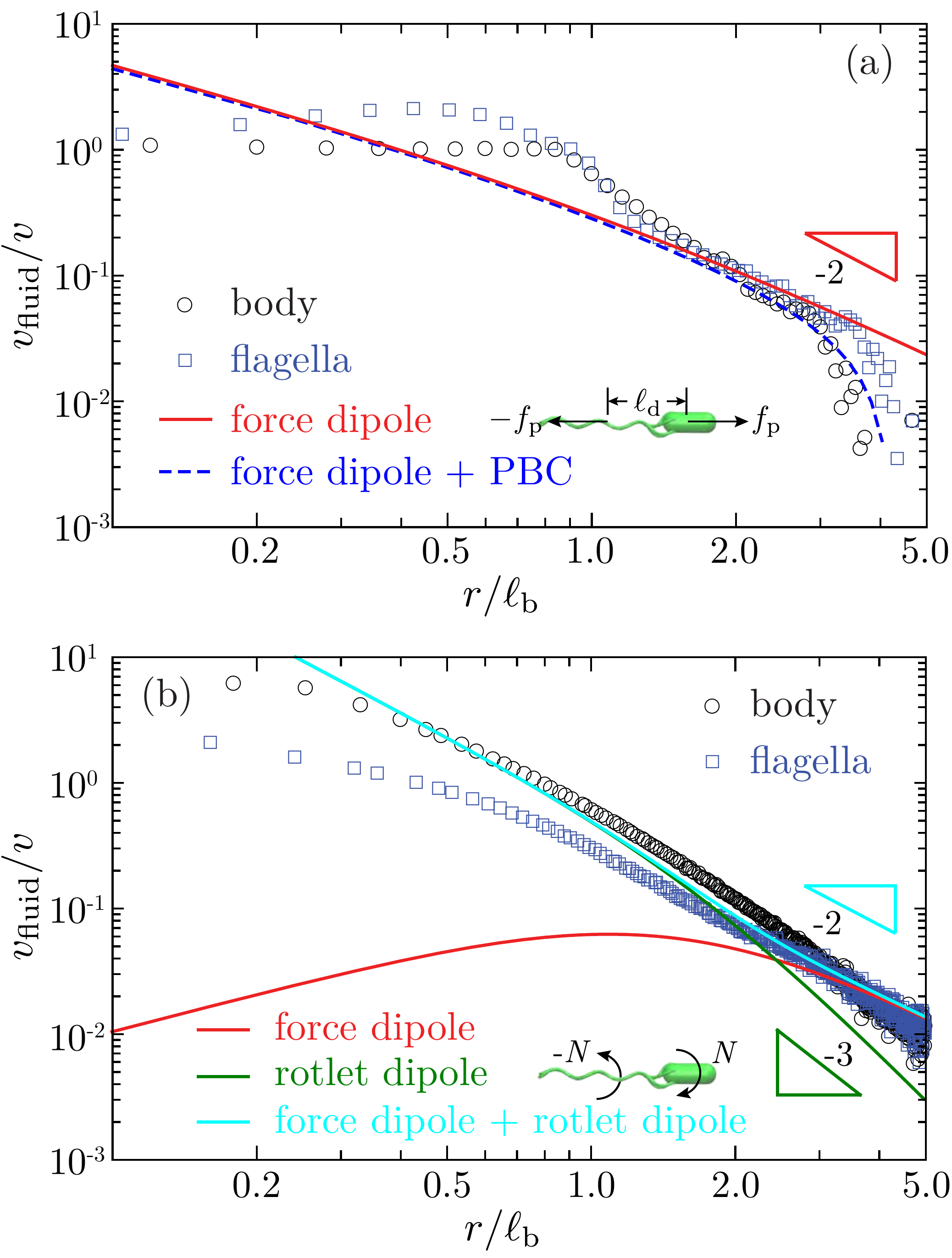}}
\caption{The fluid velocity $v_{\rm fluid}$ as a function of distance $r$ to the center of mass of the body or flagellar bundle (a) along the swimming axis, and (b) in the planes perpendicular to the swimming axis through the body or flagellar center of mass, respectively. $v$ is the swimming speed and $\ell_{\rm b}$ the body length. In (a), the solid line is the theoretical result for a force dipole of strength $f _{\rm p} \ell_{\rm d}$, where $f_{\rm p} \simeq 0.57$ pN is the bacterial propulsion force and $\ell_{\rm d} = 3.84$ \textmu m the center-of-mass separation between the body and flagellar bundle. The dashed line is for the force dipole with the same periodic boundary conditions (PBC) as in simulations. In (b), the red and green solid lines represent the results for the force dipole as in (a) and the rotlet dipole with torque $N \simeq 80 \, k_BT$, one-tenth of the torque rotating the flagellar bundle $\gamma_{\rm f}^{\rm r}\omega \simeq 800 \, k_BT$. The cyan solid line is the superposition of the force dipole and rotlet dipole.}
\label{Fig:5}
\end{figure}

We analyse next, how well the flow field can be reproduced by a simplified description in terms of a force-dipole and rotlet-dipole model. Figure \ref{Fig:5}(a) displays the fluid velocity along the swimming axis both in front of the body and behind the flagellar bundle. The theoretical result for the force dipole, with the force $f_{\rm p} \simeq 0.57$ pN and the $\ell_{\rm d} = 3.84$ \textmu m (see App.~\ref{sec:app_force-dipol}), agrees quite well with the simulation data for distance $r/\ell_{\rm b} > 1$, i.e., more than one body length away from the cell.  The flow for $r/\ell_{\rm b} > 3$ decays much faster than $r^{-2}$ due to the periodic boundary conditions. In this regime, the flow field is consistent with the result for the finite-distance force dipole of two Stokeslets with periodic boundary conditions (cf. App.~\ref{sec:app_force-dipol}). \RW{In the range
$0.8< r/\ell_{\rm b} <1.8$,  $v_{\rm fluid}$ decreases approximately as $r^{-3}$, a dependence previously observed in Ref.~\onlinecite{Larson:2010bj}. Already for $r/\ell_{\rm b} \gtrsim 2$, our flow field agrees well with that of the finite-distance force dipole. Hence,  up to $r/\ell_{\rm b} \lesssim 3 -4$ the flow field is hardly affected by the boundary conditions. The asymptotic force-dipole dependence $v_{\rm fluid} \sim r^{-2}$ is only reached for distances
$r/\ell_{\rm b} > 5$.}
Figure \ref{Fig:5}(b) displays the flow field in planes perpendicular to the swimming axis  through the center of mass of the body or flagellar bundle, respectively. The simulation results for $r/\ell_{\rm b} \gtrsim 1$ are well described by a  superposition of a force dipole with force $f_p \simeq 0.57 \mathrm{pN}$ and a rotlet dipole with the torque $N \simeq 80 \, k_BT$, one-tenth of the torque rotating the flagellar bundle $\gamma_{\rm f}^{\rm r}\omega \simeq 800 \, k_BT$. The rotlet-dipole contribution dominates at $r/\ell_{\rm b} < 2$, whereas the force-dipole contribution dominates at larger distances. This suggests that the counterrotation of cell body and flagella needs to be taken into account for an appropriate description of near-field hydrodynamics of bacteria. Our result provide the first measurement of the rotlet-dipole strength, which is important for a quantitative description of near-surface swimming behaviors of bacteria. \cite{Lauga:2014pf, Hu:2015scirep} It would be interesting to compare our predictions with future experimentally determined  rotlet-dipole strengths for flagellated bacteria.

\section{Conclusions} \label{sec:conclusion}

We have developed a coarse-grained model for an {\em E. coli}-type bacterium and investigated its swimming behavior by mesoscale hydrodynamics simulations. The suggested implementation of octahedron-type segments to construct a model flagellum of helical structure with bending and torsion elastic energy is particularly suitable in combination with a particle-based fluid, because it provides a simple means to capture the twist by the fluid on a flagellum. The hydrodynamic friction coefficients of the model bacterium and the relation between its swimming speed and flagellar rotation rate both quantitatively agree with experimental results for {\em E. coli}. The flow field created by the model bacterium exhibits a rather complex pattern adjacent to the bacterium, with and a \RW{two-vortex} structure due to the counterrotation of the cell body and flagellar bundle. From comparison to theoretical predictions for force dipole and rotlet dipole, we find the simplified dipole model can well describe the flow more than one cell-body length away from the bacterium, and the rotation of the body and flagella makes a dominant contribution to the near-field flow. We provide the first measurement of the rotlet-dipole strength, an essential quantity for modeling bacterial motion near surfaces.


The simulations show that our model is very well suited for theoretical studies of swimming bacteria. By adjusting the geometric parameters, one expects to achieve a similar quantitative description for other bacteria such as {\it Bacillus subtilis}, {\it Saolmonella typhimurium}, {\it Rhodobacter spheroides}, and {\it Rhizobium lupini}.

Naturally, further aspects of swimming bacteria can be investigated by our approach. By modifications of the elastic deformation energy in Eq. (\ref{Eq:Vel}) to account for flagellar polymorphic transformations \cite{Vogel:2010epje, Vogel:2013prl} and by implementing reversal of the flagellar motor torque ${\bf T}$, the full run-and-tumble motion can be addressed. Since boundaries and external flow are easily implemented in the MPC method, our simulation approach opens an avenue for detailed studies of confinement effects \cite{elge:10} and non-equilibrium aspects in bacteria locomotion.

\appendix

\section{Flagellum model} \label{sec:app_model}

We estimate the bending and twist strength $K_{{\rm el}}^{\alpha}$ of our discrete flagellum by mapping the elastic energy in Eq. (\ref{Eq:Vel}) to a continuous form. We identify the directions ${\bf e}_n^3$  of the bond vectors ${\bf b}_n^3$ as local tangent vectors on the filament. Then, the discrete measure of the local curvature $\kappa = \vert {\bf e}_{n+1}^3 - {\bf e}_n^3 \vert / a = 2\sin(\vartheta_n/2)/a$ turns into $\kappa \approx \vartheta_n/a$ for small bending angles $\vartheta_n$ and the local torsion becomes $\tau = - \vert {\bf n}_{n+1} - {\bf n}_n \vert / a =- 2\sin(\psi_n/2) / a \approx -\psi_n / a$ for small torsional angles $\psi_n$ between the normal vectors ${\bf n}_n$ and ${\bf n}_{n+1}$, where
${\bf n}_n = ({\bf e}_n^3 \times {\bf e}_{n+1}^3) / \vert {\bf e}_n^3 \times {\bf e}_{n+1}^3 \vert$. At small $\vartheta_n$, $\psi_n \approx \varphi_n$ since the deformation mainly originates from twist. In Eq. (\ref{Eq:Vel}), the components of the strain vector are
\begin{align}
\Omega_n^1 &= \frac{-\vartheta_n}{\sin\vartheta_n} {\bf e}_n^2 \cdot {\bf e}_{n+1}^3, \label{Eq:omega1}\\
\Omega_n^2 &= \frac{\vartheta_n}{\sin\vartheta_n} {\bf e}_n^1 \cdot {\bf e}_{n+1}^3, \label{Eq:omega2}\\
\Omega_n^3 &= \varphi_n. \label{Eq:omega3}
\end{align}
Using ${\bf e}_{n+1}^3 = \sin\vartheta_n\cos\phi_n{\bf e}_n^1 + \sin\vartheta_n\sin\phi_n{\bf e}_n^2 + \cos\vartheta_n{\bf e}_n^3$, where $\phi_n$ is the azimuthal angle of ${\bf e}_{n+1}^3$ in the plane defined by ${\bf e}_n^1$ and ${\bf e}_n^2$, expansion of Eq.~(\ref{Eq:Vel}) yields
\begin{align}
U_{\rm el} &= \frac{1}{2}\sum\limits_{n=1}^{N-1}\big\{K_{\rm el}^1\left[\vartheta_n^2+\vartheta_{\rm e}^2-2\vartheta_n\vartheta_{\rm e}\cos(\phi_n-\phi_{\rm e})\right]+
K_{\rm el}^3(\varphi_n-\varphi_{\rm e})^2\big\}\nonumber\\
&\approx \frac{1}{2}\sum\limits_{n=1}^{N-1}\left[K_{\rm el}^1(\vartheta_n-\vartheta_{\rm e})^2 + K_{\rm el}^3(\varphi_n-\varphi_{\rm e})^2\right] \nonumber\\
&\approx \frac{1}{2}\int_0^{aN}\left[K_{\rm el}^1a(\kappa-\kappa_{\rm e})^2 + K_{\rm el}^3a(\tau-\tau_{\rm e})^2\right] {\rm d}s ,
\end{align}
with the $\Omega_n^{\alpha}$ of the linearized expressions of Eqs.~(\ref{Eq:omega1})--(\ref{Eq:omega3}) and for small deviations of $\phi_n$ from its equilibrium value $\phi_e$. The integral over the flagellar contour describes the continuous elastic energy.\cite{Yamakawa:1997}

The experimental values for the bending stiffness of bacterial flagella range from about $10^{-24}$ to $10^{-21}$ N m$^2$.\cite{Fujime:1972jmb, Trachtenberg:1992jsb, Darnton:2007jb} For {\it S. typhimurium}, $K_{\rm el}^1a$ is estimated as $2-4 \times 10^{-24}$ N m$^2$ from quasi-elastic light scattering in Ref.~\citenum{Fujime:1972jmb}, consistent with the value $3.5\times 10^{-24}$ N m$^2$ measured from pulling experiments in Ref.~\citenum{Darnton:2007jb}. In the Ref.~\citenum{Trachtenberg:1992jsb}, a Young's modulus of about $10^{-11}$ N m$^{-2}$ has been determined for {\it S. typhimurium} and {\em E. coli}  from electron micrographs, which yields the approximate bending stiffness  $10^{-21}$ N m$^2$ given a filament radius of 10 nm. The twist or torsional stiffness has been generally assumed to be of the same order of magnitude as the bending stiffness. Our choice $K_{\rm el}^1=K_{\rm el}^2=K_{\rm el}^3=5\times 10^4\, k_BT$, with the physical length scale $a \approx 0.1$ \textmu m, corresponds to a bending stiffness of $K_{\rm el}^1a \simeq 2 \times 10^{-23}$ N m$^2$ well within the experimental range.

In order to determine the equilibrium parameters $\Omega_{\rm e}^\alpha$ in Eq. (\ref{Eq:Vel}), we consider a flagellum in the normal state---a three-turn left-handed helix of radius 0.2 \textmu m and pitch 2.2 \textmu m. The bending angle between subsequent discrete bond vectors is then $\vartheta_{\rm e} = \arccos{({\bf e}_n^3\cdot{\bf e}_{n+1}^3)}=0.125$ for the bond length $a=0.1$\textmu m. We choose ${\bf e}_n^1$ in each segment such that it coincides with ${\bf n}_n$ after the twist with angle $\varphi_{\rm e}$, {\it i.e.}, ${\tilde{\bf e}}_n^1 = {\bf n}_n$. ${\tilde{\bf e}}_n^1$ is thus invariant in bending from ${\bf e}_n^3$ to ${\bf e}_{n+1}^3$ around ${\bf n}_n$ by the angle $\vartheta_{\rm e}$, hence, ${\bf e}_{n+1}^1= {\tilde{\bf e}}_n^1={\bf e}_n^3\times{\bf e}_{n+1}^3/\sin\vartheta_{\rm e}$. Then, $\varphi_{\rm e} = -\arccos({\bf e}_n^1\cdot{\bf e}_{n+1}^1)=-0.217$. From Eqs. (\ref{Eq:omega1})--(\ref{Eq:omega3}), we obtain $\Omega_{\rm e}^1 = 0.122$, $\Omega_{\rm e}^2 = -0.027$, and $\Omega_{\rm e}^3 = -0.217$, which are the values used in our simulations.

\section{Force-dipole flow field for periodic boundary conditions} \label{sec:app_force-dipol}

We consider the flow generated by a force dipole as illustrated by the cartoon in Fig.~\ref{Fig:5}(a). The two point forces ${\bf f}_1=(-f_{\rm p},0,0)$ and ${\bf f}_2=(f_{\rm p},0,0)$ are located at ${\bf r}_1=(-\ell_{\rm d}/2,0,0)$ and ${\bf r}_2=(\ell_{\rm d}/2, 0,0)$, respectively. The fluid velocity at position ${\bf r}$ is given by
\begin{equation}
{\bf u}({\bf r}) = G({\bf r}-{\bf r}_1) \cdot {\bf f}_1 + G({\bf r}-{\bf r}_2) \cdot {\bf f}_2,
\end{equation}
with the Oseen tensor
\begin{equation}
G({\bf r}) = \frac{1}{8\pi \eta \vert {\bf r} \vert}\Bigl({\bf 1}+\frac{{\bf r}\otimes{\bf r}}{\vert {\bf r} \vert^2}\Bigr).
\end{equation}
To calculate the flow of the force dipole in a cubic box of length $L$ with periodic boundaries, one can start with the Oseen tensor in ${\bf k}$-space
\begin{equation}
\tilde{G}({\bf k}) = \frac{1}{\eta \vert {\bf k} \vert^2}\Bigl({\bf 1}-\frac{{\bf k}\otimes{\bf k}}{\vert {\bf k} \vert^2}\Bigr).
\end{equation}
The ${\bf k}$-space fluid velocity is then
\begin{equation}
\tilde{\bf u}({\bf k}) = \tilde{G}({\bf k}) \tilde{\bf f}_1({\bf k}) +  \tilde{G}({\bf k}) \tilde{\bf f}_2({\bf k}),
\end{equation}
with the Fourier-transformed point force $\tilde{\bf f}_i({\bf k})=\int {\bf f}_{i} {\bf \delta}({\bf x}-{\bf r}_i) e^{-i {\bf k}\cdot {\bf x}}{\rm d}^3{\bf x} =  {\bf f}_{i} e^{-i {\bf k}\cdot {\bf r}_i}$, $i=1,2$. Inverse Fourier transformation for the periodic system yields the flow field
\begin{equation}
{\bf u}'({\bf r}) = \sum_{{\bf k}\ne 0}\tilde{\bf u}({\bf k}) e^{i{\bf k} \cdot {\bf r}}
\end{equation}
with the three components of ${\bf k}$ assuming multiples of $2 \pi / L$.

\section{Resistive-force theory for helix} \label{sec:app_rft}

Expressions for the friction coefficients of Eqs.~(\ref{Eq:F}) and (\ref{Eq:T}) for a helix have been presented in Refs.~\onlinecite{ligh:76,Lauga:2006bj,Chattopadhyay:2006pnas,rode:13}. The original expressions of Ref.~\onlinecite{ligh:76} are
\begin{align} \label{eq_app_gamma}
\gamma_{\rm f}^{\rm t} = & \ K_n L_{\|} \frac{\epsilon^2 + \chi}{\sqrt{1 + \epsilon^2}} , \\
\gamma_{\rm f}^{\rm c} = & - K_n L_{\|} R \frac{\epsilon ( 1-  \chi)}{\sqrt{1 + \epsilon^2}} , \\
\gamma_{\rm f}^{\rm r} = & \ K_n L_{\|} R^2 \frac{1+ \chi \epsilon^2}{\sqrt{1 + \epsilon^2}} ,
\end{align}
with $\chi = K_t/K_n$, $\epsilon = \tan \zeta$, where $\zeta$ is the pitch angle, and
\begin{align} \label{eq_app_kn}
K_n = & \ \frac{4 \pi \eta}{\ln(c \Lambda/r) +1/2} , \\ \label{eq_app_trans}
K_t = & \ \frac{2 \pi \eta}{\ln(c \Lambda/r)} .
\end{align}
Here, $\Lambda$ is the pitch, $r$ the (hydrodynamic) radius of the flagella filament, and $c=0.18$ a constant.\cite{ligh:76} For $\chi =1/2$, which corresponds to the ratio of the viscous coefficients of an infinitely long rod, and $c=2$, the Eqs.~(\ref{eq_app_gamma}) -- (\ref{eq_app_trans}) reduce to the definitions in Ref.~\onlinecite{Lauga:2006bj}. The viscous coefficients of Ref.~\onlinecite{Chattopadhyay:2006pnas} are somewhat different, specifically the expression for $K_t$.

\section*{Acknowledgement}
We thank Adam Wysocki and Anoop Varghese for helful discussions. Financial support by the VW Foundation (Volkswagen\-Stiftung) within the program {\em Computer Simulation of Molecular and Cellular Biosystems as well as Complex Soft Matter} of the initiative {\em New Conceptual Approaches to Modeling and Simulation of Complex Systems} is gratefully acknowledged.

\balance

\footnotesize{


\providecommand*{\mcitethebibliography}{\thebibliography}
\csname @ifundefined\endcsname{endmcitethebibliography}
{\let\endmcitethebibliography\endthebibliography}{}

}

\end{document}